\documentclass{PoS}

\usepackage{natbib}

\def\vqsqusgr{\mbox{V4641 \mbox{Sgr}}}

\def\xtejodh{\mbox{XTE J1118+480}}

\def\xtejdsv{\mbox{XTE J1720-318}}

\def\cmmoinsdeux{\mbox{ cm}^{-2}}

\def\mags{\mbox{ magnitudes}}

\def\hour{^{h}}

\def\min{^{m}}

\def\secp{{\rlap.}^{s}}

\def\adeg{^{\circ}}

\def\amin{^\prime}

\def\asecp{{\rlap.}^{\prime \prime}}

\def\Av{A_{\rm v}}

\def\nh{N_{\rm H}}

\def\ltsima{\; \buildrel < \over \sim \;}
\def\simlt{\lower.5ex\hbox{\ltsima}}            % < over MMM
\def\gtsima{\; \buildrel > \over \sim \;}
\def\simgt{\lower.5ex\hbox{\gtsima}}            % > over MMM

\title{Multi-wavelength observations of the microquasar XTE J1720-318: 
a transition from high-soft to low-hard state}

\ShortTitle{Multi-wavelength observations of 2 states in the microquasar XTE J1720-318}

\author{\speaker{Sylvain Chaty}\thanks{Based on ESO observations through programme \# 070.D-0340}\\
        AIM - Astrophysique Interactions Multi-\'echelles 
(UMR 7158 CEA/CNRS/Universit\'e Paris 7 Denis Diderot), CEA Saclay,
DSM/DAPNIA/Service d'Astrophysique, B\^at. 709, L'Orme des Merisiers,
FR-91191 Gif-sur-Yvette Cedex, France.\\
        E-mail: \email{chaty@cea.fr}}

\abstract{% context heading (optional)
   {}
   {To gain a better understanding of high-energy Galactic sources,
we observed the Galactic X-ray  binary and black hole candidate $\xtejdsv$
in the optical  and  near-infrared, 
just  after the onset  of its X-ray outburst in January, 2003.
These observations were obtained with the ESO/NTT as a Target of Opportunity,
in February and April 2003.}
   {We performed an accurate astrometry and analysed photometrical 
observations. We then produced a colour-magnitude
diagram, looked at the overall evolution of the multi-wavelength light curve,
 and analysed the  spectral energy distribution.}
   {We  discovered the optical counterpart in the R-band (R $\sim 21.5$) and
confirmed the near-infrared counterpart. 
We show that, for an absorption between 6 and 8 magnitudes, 
$\xtejdsv$ is  likely to be an intermediate  mass X-ray binary 
located  at a distance  between 3  and  10 kpc,  hosting  a  main
sequence star  of spectral type between  late B and early  G.
Our second set of observations took place simultaneously with the third
 secondary outburst present in X-ray and near-infrared light curves.
 The evolution  of its spectral  energy distribution 
shows that $\xtejdsv$ entered a transition from a high-soft to a low-hard
state in-between the two observations.}
   {We finally discuss the different phases of the 
outburst of this source in the general scheme of outbursts from
microquasars.}}

\FullConference{VI Microquasar Workshop: Microquasars and Beyond\\
		September 18-22 2006\\
		Societ\`a del Casino, Como, Italy}

\begin{document}

\section{Discovery and characteristics of $\xtejdsv$}

X-ray binaries are constituted of a compact object and a companion
star, the former attracting matter from the later, either through
an accretion disc or the wind. They are usually divided in 2 sub-classes:
high mass X-ray binaries and low mass X-ray binaries, hosting early-type
and late-type stars, respectively. Since accretion and ejection phenomena
usually occur in these objects, they are ideal laboratories
for studying relativistic phenomena and the formation
and evolution of compact objects in binaries.
However, to study them, we first have to derive 
the important parameters related to
the nature of these systems, i.e. the distance, nature of the compact object, 
spectral type of the companion star, type of accretion, orbital parameters, 
etc.  Because of the way
they are formed, most of the observed Galactic X-ray binaries are located
in the Galactic plane or even towards the Galactic centre, 
and therefore associated with very high
absorption (up to $\Av \sim 50 \mags$) 
because of the presence of gas and dust
in this region. In this case, near-infrared (NIR) observations
prove to be particularly useful, since the radiation is less absorbed
at NIR wavelengths than at optical ones (see, e.g., \cite{chaty:2002b}).
Furthermore,  X-ray binaries have to be studied in a multi-wavelength context
to disentangle all parts of the system emitting at various 
wavelengths: the accretion disc from high-energies to NIR, 
the companion star from
ultra-violet to NIR, the jets from radio to X-rays, etc. 
(see, e.g., \cite{chaty:2006a}).

On January 9, 2003, the All Sky Monitor ({\it ASM}) of the {\it
Rossi-XTE} satellite discovered a new source in the X-ray sky:
$\xtejdsv$, in the direction of the Galactic bulge, at $6 \adeg$ from
the Galactic Centre.  The 2-12 keV flux was initially $\sim$
130 mCrab and reached $\sim$ 430 mCrab in one day, on January 10,
2003 \cite{remillard:2003}. 
Spectroscopic observations with {\it XMM-Newton} were carried
out on February 20, 2003, allowing \cite{markwardt:2003a}
 to estimate the column density of
hydrogen on the line of sight: $\nh=1.33 \times 10^{22}
\cmmoinsdeux$. An iron line was detected at 6.2 keV with 95 eV
equivalent width, and no low or high frequency oscillation was
detected \cite{markwardt:2003a}.
The 2-10 keV flux was estimated to be $1.6 \times 10^{-9} \mbox{ erg
cm}^{-2} \mbox{s}^{-1}$ \cite{gonzalez-riestra:2003}.  
The source was not detected during
{\it INTEGRAL}/{\it IBIS} observations on February 28, 2003, but
became visible at the end of the outburst during {\it IBIS} surveys of
the Galactic centre from the end of March 2003
\cite{goldoni:2003}. The source
reached $\sim$ 25 mCrab in the 15-40 keV band and became detected in
the 40-100 keV energy band at $\sim$ 30 mCrab on April 6 and 7, 2003,
indicating that the source had undergone a change of state, as
suggested by \cite{goldoni:2003}.  The
high-energy observations suggest that the compact object is a black
hole, first because of its early spectral evolution that was very
similar to black hole X-ray transients \cite{remillard:2003}, 
and also because of X-ray spectral
parameters \cite{cadolle-bel:2004}
and the presence of an iron line \cite{markwardt:2003a}.
A radio counterpart was discovered with the VLA on January 15, 2003, and
confirmed with ATCA on January 16, 2003: only one
radio source included in the {\it Rossi-XTE} error box significantly varied 
from 0.32 $\pm$ 0.04 to 4.9 $\pm$ 0.1 mJy at 4.9 GHz \cite{rupen:2003}.  
Radio observations took place from January to August 2003, allowing 
\cite{brocksopp:2005} to study
the correlation between radio and X-ray fluxes.  
A NIR counterpart was discovered then by \cite{kato:2003} 
on January 18, 2003. Thirteen observations in
$J$, $H$, and $K_{s}$, until May 21, 2003, allowed \cite{nagata:2003}
 to measure the
exponential decay following the outburst, which was equal to 60 days.

In this paper, we will first describe our optical and NIR observations
and data reduction in Sect. \ref{obs}, then report on our astrometry and 
photometry results in Sect. \ref{results}.
We will then focus on constraining the companion star spectral
type in Sect. \ref{companion}, and finally we will analyse the evolution
of the $\xtejdsv$ light curve and 
spectral energy distribution (SED) in Sect. \ref{SED}.
Readers who want to have more details might refer to \cite{chaty:2006b}.

%__________________________________________________________________

\section{ESO Optical/NIR observations}

     \subsection{Observations and data analysis} \label{obs}

Our observations were carried out as part of the Target of Opportunity
(ToO) programme 070.D-0340 (PI: S. Chaty) dedicated to the study of
new Galactic high-energy sources and jet sources. They were triggered
so as to be conducted at the same time as {\it INTEGRAL} ToO
observations, and we asked for two periods of observations.  The first
set of observations took place on February 28, 2003, and the second
one on April 24, 2003.  On February 28, 2003, we obtained NIR
photometry in $J$-, $H$-, and $K_{s}$-bands with the spectra-imager
SofI, and optical photometry in $B$-, $V$-, $R$-, and $I$-bands with
EMMI, both installed on the NTT (La Silla Observatory, Chile, European
Southern Observatory). We used the large field imaging of SofI's
detector, giving an image scale of 0.288" pixel$^{-1}$ and a field of
view of $4.94\amin \times 4.94\amin$, and the EMMI detector with an
image scale of 0.32" pixel$^{-1}$ and a binning $2 \times 2$, giving a
field of view of $9.9\amin \times 9.1\amin$.
Concerning the NIR observations, we repeated one set of observations
for each filter with 9 different 30" offset positions, including
$\xtejdsv$, with an integration time of 90 seconds for each exposure,
following the standard jitter procedure that allows us to cleanly
subtract the blank NIR sky emission. We observed two photometrical
standard stars of the faint NIR standard star catalogue of
\cite{persson:1998}: sj9157 on
February 28, 2003, and sj9172 on April 24, 2003. Concerning the
optical observations, we acquired 300 s exposures in each filter,
except for the B-band (200 s), using a 2x2 binning to increase the
sensitivity. We observed the standard star RU152 in R- and I-bands.
Since we did not have any standard star observations in the B and V
filters, we used mean zero-points taken from the EMMI
website\footnotemark[1].
\footnotetext[1]{www.ls.eso.org/lasilla/sciops/ntt/emmi/} 
We used the Image Reduction and Analysis Facility \rm{(IRAF)} suite to
perform data reduction, carrying out standard procedures of optical
and NIR image reduction, including flat-fielding and NIR sky
subtraction.  As we had only one standard star observation
available for each night, we used characteristic extinction
coefficients at la Silla: $ext_B=0.214$, $ext_V = 0.125$,
$ext_R=0.091$, $ext_I= 0.051$, $ext_J = 0.08$, $ext_H = 0.03$, and
$ext_{Ks} = 0.05$, to transform instrumental magnitudes into apparent
magnitudes.  The observations were performed through an airmass
between 1 and 1.4.

     \subsection{Astrometry and photometry results} \label{results}

We used the $K_{s}$ image of the $\xtejdsv$ field taken on  January 21, 2003
\cite{obrien:2003}
to identify $\xtejdsv$ in our NTT images. We then determined 
the position of the $\xtejdsv$ NIR counterpart by deriving the astrometric
solution, using $\sim 12$ stars taken from the GSC2 catalogue: the position we
measured was: $\alpha = 17\hour 19\min 58\secp 988 \pm0\secp 008$;
$\delta = -31\adeg 45\amin 01\asecp 21 \pm 0\asecp15$ (equinox
J2000). This position is consistent with other determinations
(Table \ref{astrometry}).
We discovered the optical counterpart in the R- and I-bands at $\alpha
= 17\hour 19\min 58\secp994 \pm 0\secp007$, $\delta = -31\adeg 45\amin
01\asecp46 \pm 0\asecp15$ (equinox J2000), a position that is consistent 
with the NIR counterpart. 
We present $BVRI$ magnitudes in Table \ref{mag_optir}. The R and I 
magnitudes are consistent with detection
limits given by \cite{nagata:2003}: $R>18$ and $I>16.5$. 
We give a lower limit for the B- and V-bands, as we did not detect
any counterpart in these bands.
Since $\xtejdsv$ is located close to the Galactic centre, we had to
perform crowded field photometry to obtain precise NIR
magnitudes, using the {\it noao.daophot} package. This procedure,
described in \cite{massey:1992}, consists of creating an empirical
point-spread function with isolated bright stars, applying this model to the
whole field, cancelling the contributions of neighbour stars, measuring
the flux of the object itself, and then applying aperture correction
(due to the use of a smaller aperture for measuring $\xtejdsv$ magnitude
than for standard stars). 
This procedure allows us to get photometry with better than 1\% accuracy.  
We present
the apparent $J$, $H$, and $K_{s}$ magnitudes measured in  February and
April 2003 in Table \ref{mag_optir}. Uncertainties were determined from CCD
readout and signal noise. We point out that we do not include  
the $H$-band observation of  February 28, 2003, 
here because of bad sky subtraction.
 
\begin{figure}
\resizebox{\hsize}{!}{\includegraphics[width=6cm, angle=0]{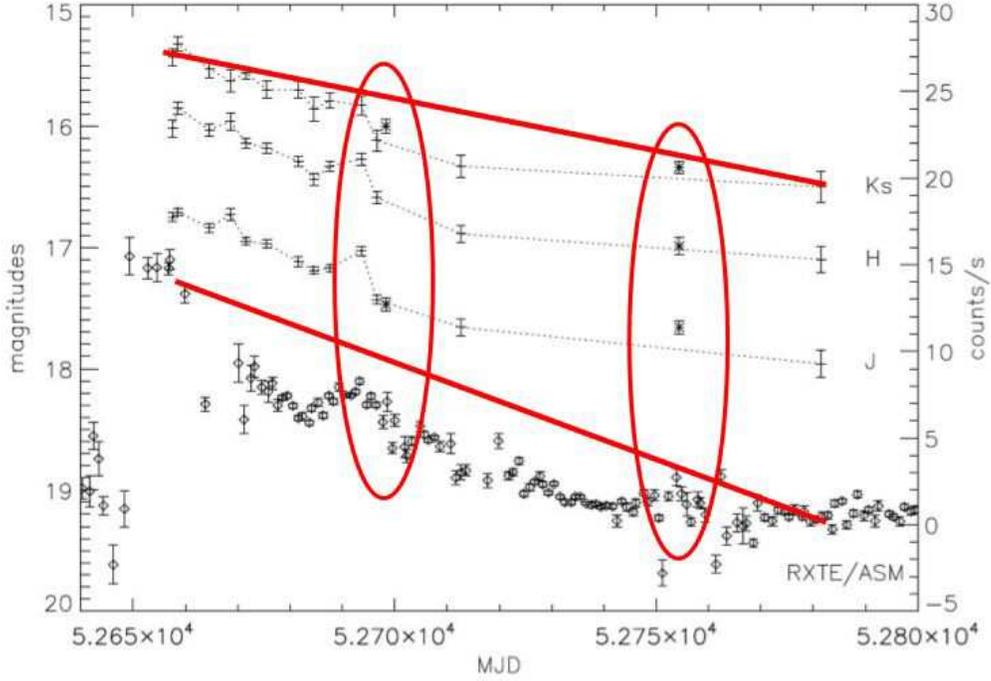}}
\caption{Multi-wavelength $\xtejdsv$ light curve.
From bottom to top: {\it Rossi-XTE} X-ray light curve (indicated
by diamonds); NIR J, H, and $K_{s}$ light curves, respectively.
NIR data taken from \cite{nagata:2003} are reported with '+', 
and data from this paper with '$\ast$' (MJD = JD - 2400000.5).
The red lines indicate that the NIR flux decayed less rapidly
than the X-ray flux, in agreement with the Disc Instability Model. 
The red ellipses indicate the time of our observations, allowing 
to reveal two simultaneous
X-ray and NIR outbursts.}
\label{IRlc}
\end{figure}

Our NIR photometrical observations are reported in Fig. \ref{IRlc},
where we also included results from \cite{nagata:2003}. 
Our magnitudes, indicated by '$\ast$', are
consistent with measures from \cite{nagata:2003}.  
We notice the similar behaviour of X-ray
and NIR light curves, particularly during the first maximum when the
coverage is more complete. Although similar, the NIR flux decayed
less rapidly than the X-ray flux, as indicated by the red lines in Figure
\ref{IRlc}. This is a consequence of the 
accretion-disc instability model, where the optical and NIR emission
emanates from the outer part of the accretion disc, which is kept
hot enough long until the cooling front from inside reaches the outer
part \citep{lasota:2001}.
We point out that we observed an
increase in NIR on April 24, 2003 (MJD 52753), observed by chance at
exactly the same time as an X-ray increase seen on the light curve.
These ESO observations are therefore simultaneous to the 2 outbursts,
as indicated by the red ellipses in Figure \ref{IRlc}.

%%%%%%%%%%%%%%%%%%%%%%%%%%%%%%%%%%%%%%%%%%%%%%%%%%%%%%%%%%
\begin{table*}
\centering
\begin{tabular}{cccc}
\hline \hline
  & This paper & \cite{nagata:2003} & \cite{obrien:2003} \\
\hline 
$\alpha$ & $17\hour 19\min 58\secp988 \pm 0\secp008$ & $17\hour 19\min 59\secp000 \pm 0\secp 014$ & $17\hour 19\min 58\secp 994 \pm 0\secp004$ \\
$\delta$ & $-31\adeg 45\amin 01\asecp21 \pm0\asecp15$  & $-31\adeg 45\amin 01\asecp 2 \pm 0\asecp2$ & $-31\adeg 45\amin 01\asecp 25 \pm 0.05$ \\
\hline
\end{tabular}
\caption{Summary of astrometry results (equinox J2000) for the $\xtejdsv$
NIR counterpart.}
\label{astrometry}
\end{table*}
%%%%%%%%%%%%%%%%%%%%%%%%%%%%%%%%%%%%%%%%%%%%%%%%%%%%%%%%%%

%%%%%%%%%%%%%%%%%%%%%%%%%%%%%%%%%%%%%%%%%%%%%%%%%%%%%%%%%%
\begin{table*}
\centering
\begin{tabular}{ccccccc}
\hline \hline
Date & MJD & R & I & J & H & $K_{s}$ \\
\hline 
2003 02 28 & 52698 & & & $17.47\pm0.05$ & - & $16.00\pm0.06$ \\
2003 04 24 & 52753 & $21.5\pm0.3$ & $20.6\pm0.1$ & $17.66\pm0.05$ & $16.99\pm0.07$ & $16.34\pm0.05$ \\
\hline
\end{tabular}
\caption{Apparent R, I, J, H, and $K_{s}$ 
magnitudes of $\xtejdsv$ (MJD = JD - 2400000.5).
$\xtejdsv$ is not detected in B ($>23.2\pm0.4$) or V ($>23.1\pm0.4$). 
\label{mag_optir}}
\end{table*}
%%%%%%%%%%%%%%%%%%%%%%%%%%%%%%%%%%%%%%%%%%%%%%%%%%%%%%%%%%

%%%%%%%%%%%%%%%%%%%%%%%%%%%%%%%%%%%%%%%%%%%%%%%%%%%%%%%%%%
%%%%%%%%%%%%%%%%%%%%%%%%%%%%%%%%%%%%%%%%%%%%%%%%%%%%%%%%%%

\section{The nature of the binary system and companion star} \label{companion}

\begin{figure}
\centering
\includegraphics[width=13cm,angle=0]{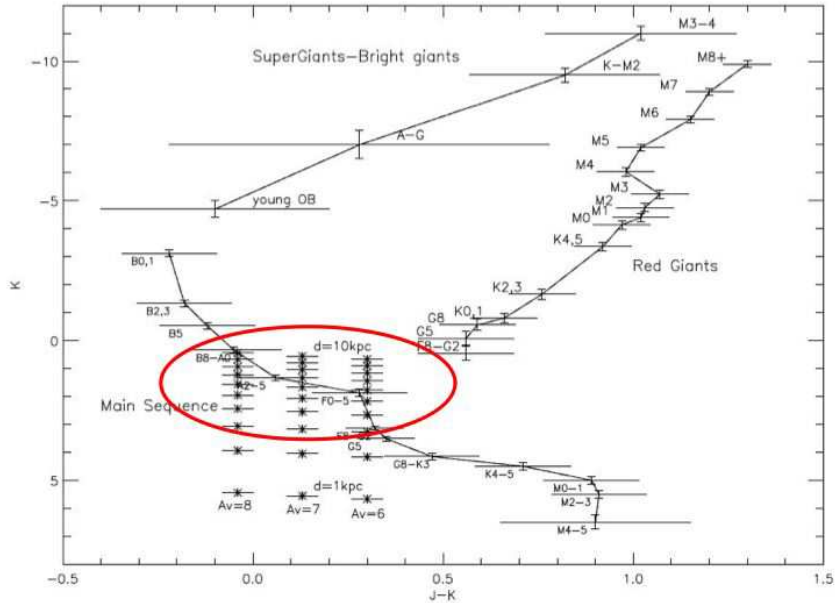} 
\caption{Colour-magnitude diagram showing characteristic absolute
magnitudes of various spectral types and $\xtejdsv$ absolute
magnitudes computed for interstellar absorption $\Av$ between 6 and
$8 \mags$ ('$\ast$' from right to left) and distance between 1 and
10 kpc ('$\ast$' from bottom to top). We used the  April 24, 2003,
observations, when the object was fainter, to reduce the
accretion disc contribution in the NIR flux. 
From this CMD we can conclude that $\xtejdsv$ is an 
intermediate mass X-ray binary located at a distance
between 3 and 10 kpc, and that the companion star is a main sequence star
of spectral type between late B and early G (see text for more details).
The ellipse indicates the favoured region.
}
\label{CMD}
\end{figure}

We will now try to constrain the nature of the companion star 
with a colour-magnitude diagram (CMD).
$\xtejdsv$ optical and NIR magnitudes allow us to constrain the nature
of the binary system by comparing its absolute magnitudes with those
of well-determined spectral type stars (see, e.g.,
\cite{chaty:2002b}). For this purpose, we
use template absolute magnitudes related to spectral types (taken from
\cite{ruelas-mayorga:1991}).  
The conversion of apparent magnitudes $m$ to absolute
magnitudes $M$ depends on both distance $d$ and interstellar
absorption $A_{v}$, via: $M=m+5-5log d(pc)-A_{v}$.  Concerning the
interstellar absorption, we have three different estimates: First,
{\it XMM-Newton} spectroscopy obtained in February 2003 gave $\nh =
1.24 \pm 0.02 \times 10^{22} \cmmoinsdeux$
\cite{cadolle-bel:2004}. This
column density corresponds to an absorption of $A_{v}=6.9 \mags$ using
the relation $\Av=5.59 \times 10^{-22} \nh$ \cite{predehl:1995}. Second, \cite{nagata:2003}  obtained $A_{v}= 8$ by assuming a high
temperature blackbody emission just after the X-ray outburst. They also
noted that extinction derived from the 2MASS survey is $A_{v} \sim 6$.
Here, we will use only one observing epoch, on  April 24, 2003,
when the source is fainter, to minimise the accretion disc
contribution in the observed NIR flux. Even if the object was still
far from quiescence, in this way, we determine a lower limit 
for the companion star spectral type 
by assuming that the accretion disc emission reddens the NIR flux.
To derive the possible spectral types, we computed the absolute
magnitudes of $\xtejdsv$, taking various distances $d$ and absorption in
the visible $A_{v}$. The results are reported with '$\ast$' 
in the ($J-Ks$, $Ks$) colour-magnitude diagram (CMD) presented in Fig. 
\ref{CMD}. The distance was computed between 1 and 10 kpc 
(from bottom to top, respectively) and the absorption between 6 and 8 magnitudes
(from right to left, respectively).

From this CMD, we first derive that, in any case, the companion star must
 belong to the main sequence, the favoured region being indicated
by the red ellipse.  Furthermore, if the interstellar
absorption is high, $\Av \sim 8 \mags$, the spectral type would be
between late B and early A, and the source far away, between 6 and 10
kpc.  With an intermediate value of the interstellar absorption,
$\Av \sim 7 \mags$, the spectral type would be between late A and
early F, and the distance between 5 and 7 kpc.  Finally, with a small
interstellar absorption, $\Av \sim 6 \mags$, the spectral type would
be between late F to early G, and the distance between 3 and 6 kpc.
Therefore, we can conclude that, for an absorption between
6 and 8 magnitudes, the $\xtejdsv$ companion star is a main sequence star
of spectral type between late B and early G, located at a distance
between 3 and 10 kpc. This estimate of distance makes the source
 closer than suggested by \cite{nagata:2003}: 
it is therefore possible that the source is not located in the Galactic bulge.
We point out that from this analysis, 
$\xtejdsv$ can be added to the list of intermediate mass X-ray binaries, 
like, e.g., $\vqsqusgr$ \cite{chaty:2003a}.

%%%%%%%%%%%%%%%%%%%%%%%%%%%%%%%%%%%%%%%%%%%%%%%%%%%%%%%%%%
%%%%%%%%%%%%%%%%%%%%%%%%%%%%%%%%%%%%%%%%%%%%%%%%%%%%%%%%%%

\section{Transition from high-soft to low-hard state} \label{SED}

\begin{figure}
\centering
\includegraphics[width=13cm,angle=0]{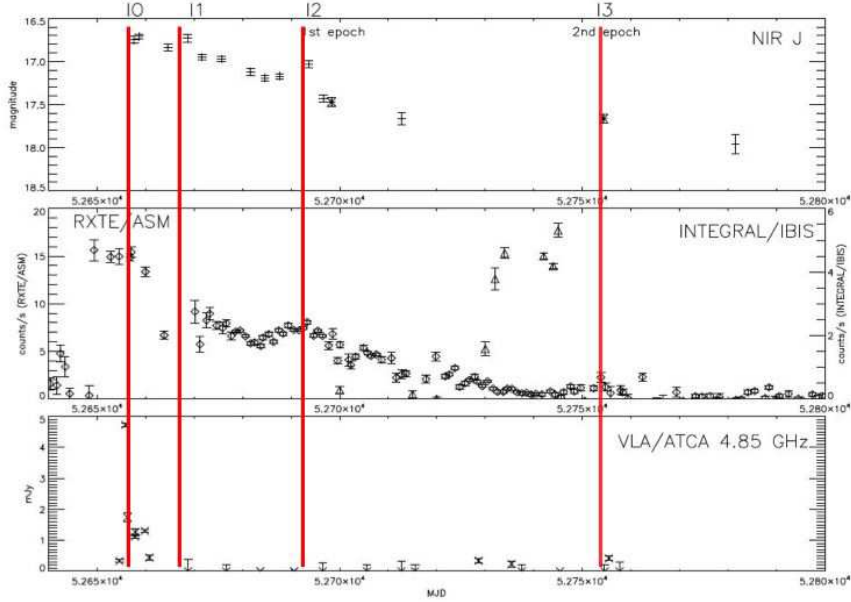}
\caption{Multi-wavelength light curve showing the outburst of $\xtejdsv$
and its transition from high-soft to low-hard state.
Top panel: NIR observations ('+': \cite{nagata:2003}, '$\ast$': this paper);
Middle panel: high-energy observations 
('$\diamond$': {\it RXTE/ASM}, '$\triangle$': {\it INTEGRAL/IBIS}, \cite{cadolle-bel:2004});
Bottom panel: radio VLA/ATCA observations ('x': \cite{brocksopp:2005}).
We indicated at the top the times when the main outburst (O)
and the three secondary outbursts (1, 2, 3 respectively) occurred,
indicated with red lines.
MJD = JD - 2400000.5}
\label{lc}
\end{figure}

In Fig. \ref{lc}, we show the overall outburst light curve of $\xtejdsv$ from 
mid-January to the end of August 2003.
First we can see that the light curve  grossly has the form
of a Fast-Rise Exponential-Decay
light curve (so-called FRED) with a timescale of 60 days, 
but superimposed on this FRED, the source 
exhibits a complex behaviour:
after the main outburst on January 16 (MJD 52656), 
we can see both in X rays and in NIR 
a secondary outburst on  January 29 (MJD 52669), 
then a second one on February 22 (MJD 52693),
and finally a third one on April 24, 2003 (MJD 52753), 
exactly at the time of our second epoch NIR observations.
The main outburst and the last event are also associated
with radio outbursts, and therefore with ejection events.
We indicate the time of these events in Fig. \ref{lc}
by O, 1, 2, and 3 in the top, respectively, and by the red lines.
There are other sources that exhibit clear secondary maxima in their
X-ray light curves, such as A0620-00, GS 1124-68, GRO J0422+32 
\cite{chen:1997}, 4U 1543-47 \cite{buxton:2004}, 
XTE J1550-564 \cite{jain:2001}
and XTE J1859+226 \citep{brocksopp:2002}.
However, $\xtejdsv$ seems to be the second source after A0620-00
to exhibit clear multi-secondary maxima in the optical/NIR,
correlated with the X rays, as seen in Fig. \ref{lc}.
This complex lightcurve and presence of multi-secondary outbursts
might be explained by the accretion-disc instability model 
(such as an ``outside-in''-type outburst).

We also report the SED of $\xtejdsv$ in Figures \ref{sedHS} and 
\ref{sedLH} for the two
observing epochs 
('+' for  February 28, 2003, and '$\ast$' for  April 24, 2003, respectively), 
where we put together quasi-simultaneous
ESO/NTT optical/NIR observations from this paper, 
{\it INTEGRAL/IBIS} high-energy observations \cite{cadolle-bel:2004}), 
and ATCA/VLA radio data \cite{brocksopp:2005}. 
We will now describe the light curve and both SEDs and analyse $\xtejdsv$'s
evolution between these two observing epochs.

%%%%%%%%%%%%%%%%%%%%%%%%%%%%%%%%%%%%%%%%%%%%%%%%%%%%%%%%%%%%%%%%%%%%%
%%%%%%%%%%%%%%%%%%%%%%%%%%%%%%%%%%%%%%%%%%%%%%%%%%%%%%%%%%%%%%%%%%%%%

	\subsection{The  February 2003 epoch: high-soft state}

\begin{figure}
\centering
\includegraphics[width=13cm,angle=0]{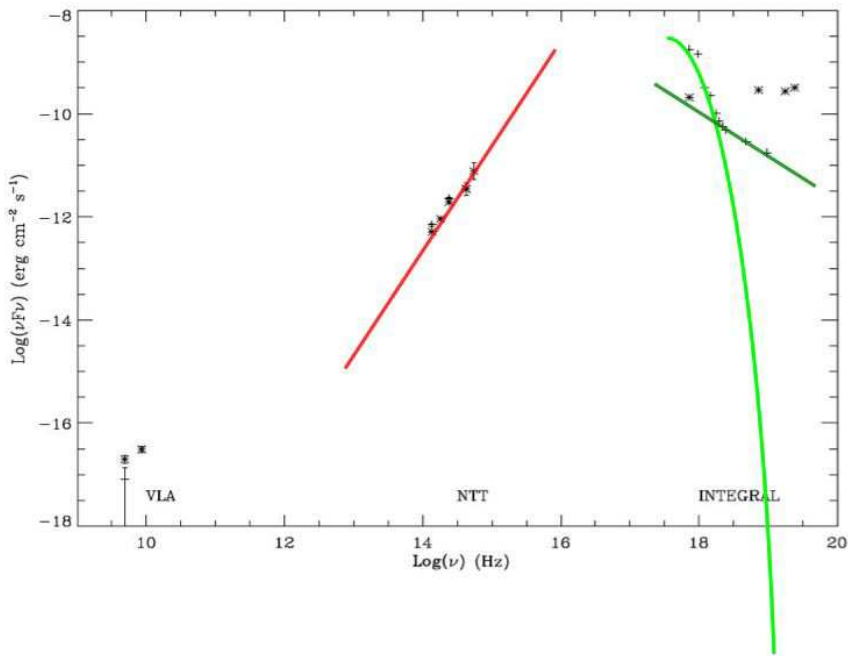}
\caption{Spectral Energy Distribution of $\xtejdsv$ during
the two observing epochs: the  February 28, 2003, data are represented by
'+', and the  April 24, 2003, by '$\ast$'.  The observations were
taken nearly simultaneously with the VLA in the radio  \cite{brocksopp:2005}, 
the NTT/EMMI and SOFI in the optical and NIR (this paper), 
and {\it INTEGRAL/IBIS} in the high-energy \cite{cadolle-bel:2004}. For
the first epoch, the VLA data were taken on  February 26, 2003, NTT/SOFI
data on  February 28, 2003, and {\it INTEGRAL/IBIS} data on  February 28
-- March 2, 2003. For the second epoch, VLA data were taken on  April
26, 2003, NTT/SOFI data on  April 24, 2003, NTT/EMMI on April 27, 2003, and
{\it INTEGRAL/IBIS} data on  February 28 -- March 2, 2003. The 
February 28, 2003, observations correspond to the high-soft state: high and
soft X-ray flux and no radio emission. On the contrary, the 
April 24, 2003, observations correspond to the low-hard state: low and hard
X-ray flux and detection of radio emission. Optical and
NIR fluxes were de-reddened, assuming an interstellar absorption in the
visible of $\Av = 7 \mags$. It is remarkable that they remain the same 
in both states. We overplot the High Soft state in this Figure:
in green the multi-colour black-body disc emission
and the soft power law in X-rays, and in red the emission in optical and NIR.
}
\label{sedHS}
\end{figure}

Observations obtained with {\it XMM-Newton}, {\it INTEGRAL/IBIS}, and 
{\it Rossi-XTE} in  February 2003 showed that the source spectrum was very
soft (power law photon index of $\Gamma = 2.7$): 
while the 2-12 keV flux was $\sim 100$ mCrab, {\it IBIS}
detected a source at only $\sim 2.1$ mCrab in the 20-120 keV band 
 \cite{cadolle-bel:2004}.
$\xtejdsv$ was not detected in radio 
($\leq 0.27$ mJy at 4.8 GHz, \cite{brocksopp:2005}),
which is consistent with typical high-soft state.
This is corroborated by the analysis of $\xtejdsv$ SED 
from  February 28, to March 2, 2003 (reported 
with '+' in Fig. \ref{sedHS}), where we notice the characteristic shape 
of the accretion disc emission in X-rays
(strong flux and soft spectrum), and the absence of radio emission.
We overplot in green the multi-colour black-body disc emission
and the soft power law in X-rays, and in red the emission in optical and NIR.

	\subsection{The  April 2003 epoch: low-hard state}

\begin{figure}
\centering
\includegraphics[width=13cm,angle=0]{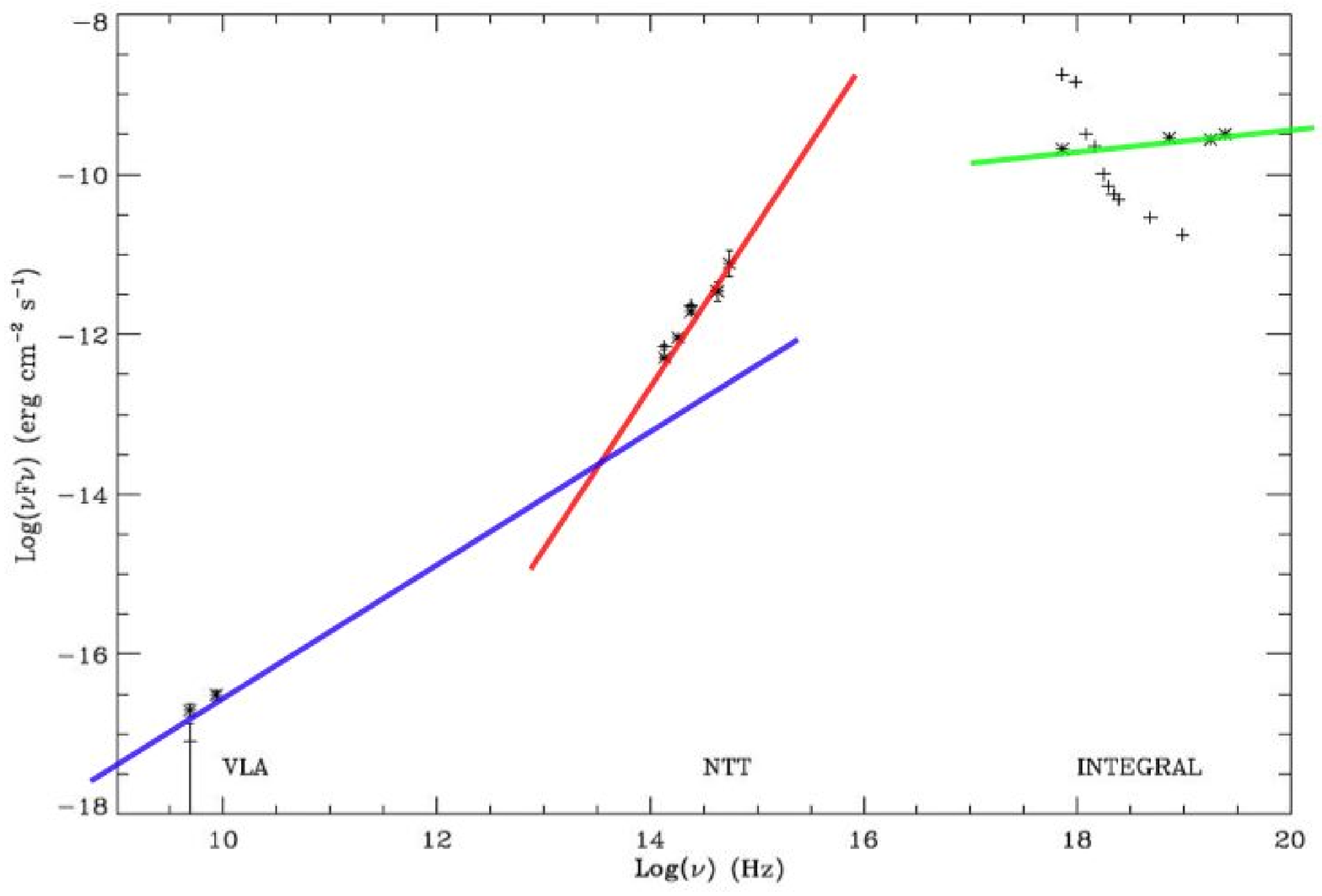}
\caption{Same Figure as above, with the Low Hard State overplot: in green 
the hard power law in X-rays, in red the emission in optical and NIR
and in blue the power law in radio.}
\label{sedLH}
\end{figure}

On the other hand, after March 25, 2003 (MJD 52723), $\xtejdsv$ hard
X-ray flux as seen by {\it INTEGRAL/IBIS} increased by a factor of 100
with respect to the high-soft state \cite{cadolle-bel:2004}, 
while the soft X-ray flux ({\it
RXTE/ASM}) remained constant, as seen on the light curve of
Fig. \ref{lc}.  Besides, we also observed a radio outburst
\cite{brocksopp:2005}, simultaneous
with an X-ray outburst and an increase in the NIR flux.  Therefore,
$\xtejdsv$ seems to have entered a transition towards a low-hard state
in-between these 2 observing epochs, as suggested by
\cite{goldoni:2003}.  This is
confirmed by the analysis of the SED (data of this second epoch are
reported with '$\ast$' in Fig. \ref{sedLH}, where we overplot in green
the hard power law in X rays, in red the emission in optical and NIR
and in blue the power law in radio): the source shows all the usual
signs of the low-hard state (see, e.g., \cite{chaty:2003b}).  
Firstly, we immediately notice that the
source hardened in the high-energy domain with a high-energy power law
photon index of $\Gamma = 1.8$.  Secondly, the radio emission is usually
interpreted in this state as synchrotron emission emanating from a
jet.  From the SED, we can derive the power law  index $\alpha$ (in
$S_\nu \propto \nu^\alpha$): $\sim -0.3$ in the radio and $\sim 1.6$
in the optical/NIR.  Therefore, the extrapolation of the radio flux
towards the optical/NIR domain is significantly fainter than the
observed optical/NIR flux.  This strongly suggests that the
synchrotron emission from the jet is contributing only for a small
part, if any, in the NIR emission.  It seems also likely that the
optical/NIR emission comes from the thermal accretion disc and the
irradiated companion star, since the slope in the NIR and optical
remarkably remains the same in both observing epochs, while the X-ray
decreased at the same time. Therefore, the NIR emission is dominated
by the contribution of the companion star, which is consistent with
$\xtejdsv$ being an intermediate mass X-ray binary.

%%%%%%%%%%%%%%%%%%%%%%%%%%%%%%%%%%%%%%%%%%%%%%%%%%%%%%%%%%%%%%%%%%%%%
%%%%%%%%%%%%%%%%%%%%%%%%%%%%%%%%%%%%%%%%%%%%%%%%%%%%%%%%%%%%%%%%%%%%%

%%%%%%%%%%%%%%%%%%%%%%%%%%%%%%%%%%%%%%%%%%%%%%%%%%%%%%%%%%%%%%%%%%%%%
%%%%%%%%%%%%%%%%%%%%%%%%%%%%%%%%%%%%%%%%%%%%%%%%%%%%%%%%%%%%%%%%%%%%%

     \subsection{Trying to put everything together}

From the analysis of the outburst lightcurve and the SED of $\xtejdsv$
we therefore see that the source exhibited the first part of its outburst
in the high-soft state, then transited to the low-hard state.
The source probably began its outburst in the low-hard state,
as tentatively 
suggested by spectral analysis of X-ray data \cite{brocksopp:2005}.
It might even have passed through the intermediate state while
it was crossing the jet line, just before the transition to the low-hard
state (see e.g. \cite{fender:2004}).

We can also point out that $\xtejdsv$ does not follow the radio--X-ray
luminosity correlation of black hole candidates in the
low-hard state (see e.g. \cite{gallo:2003}). 
Indeed, it lies below this correlation even if put at a distance 
of 1 kpc. But $\xtejdsv$ is
not the first source to deviate from this not-so-universal correlation
(see for instance SWIFT J1753.5-0127 in \cite{cadolle-bel:2006}).

%
%______________________________________________________________

\section{Conclusions}

We have reported ESO/NTT optical and NIR observations of the
microquasar $\xtejdsv$, taken as Target of Opportunity observations
following the January 2003 outburst of this source.  By performing
accurate astrometry, we discovered the optical counterpart in the
R-band (R $\sim 21.5$) and confirmed the near-infrared counterpart.
From photometric observations, analysis of a colour-magnitude diagram,
and a basic modelling of its SED, we found that, for an absorption
between 6 and 8 magnitudes, $\xtejdsv$ is likely to be an intermediate
mass X-ray binary, hosting a black hole and a main sequence star of
spectral type between late B and early G, located at a distance
between 3 and 10 kpc.  We also analysed the $\xtejdsv$ X-ray and
near-infrared light curves: superimposed to a main FRED lightcurve, it
exhibited three secondary outbursts, and our second set of
observations took place simultaneously with the third one. The study
of such outburst lightcurves is of prime importance to test accretion
disc models.  Comparing the SEDs during and after its outburst, we
confirm the change of state of this source, from high-soft to low-hard
state.

To further understand this source, it would
be useful i) to get spectroscopic observations of $\xtejdsv$ in
quiescence to better characterise the
companion star when the photospheric flux of the star dominates
and ii) to observe the radial velocity of the binary system to
derive the mass function and orbital parameters. With these 
parameters we will be able 
to further analyse its SED (as in, e.g., \cite{chaty:2003b}
with the source $\xtejodh$).
And obviously, we need some more continuous multi-wavelength coverage
of such outbursts to understand how they work.

%\bibliographystyle{/Users/chaty/Library/Texmf/Bibtex/PoS} 
%\bibliographystyle{/Users/chaty/Library/Texmf/Bibtex/amsplain} 
%\bibliographystyle{PoS} 
%\bibliography{/Users/chaty/Library/Texmf/Science/science}

\end{document}